\documentclass[lettersize,journal]{IEEEtran}
\usepackage{amsmath,amsfonts}
\usepackage{algorithmic}
\usepackage{algorithm}
\usepackage{array}
\usepackage[caption=false,font=normalsize,labelfont=sf,textfont=sf]{subfig}
\usepackage{textcomp}
\usepackage{stfloats}
\usepackage{url}
\usepackage{verbatim}
\usepackage{xcolor}
\usepackage{graphicx}
\usepackage{cite}
\usepackage[
    caption=false,
    font=normalsize,
    labelfont=sf,
    textfont=sf
]{subfig}
\begin{document}

\title{Spatially Reconfigurable Pinching-Antenna Systems: Experimental Validation and ISAC Applications}

 \author{
\IEEEauthorblockN{Shaokang Hu,
Ruotong Zhao,
Qigejian Wang,
Shaghik Atakaramians,
Derrick Wing Kwan Ng,
and Jinhong Yuan}
\IEEEauthorblockA{School of Electrical Engineering and Telecommunications, 
The University of New South Wales, Sydney, NSW 2052, Australia}  
}


\maketitle
\begin{abstract}
Future integrated sensing and communication (ISAC) networks require wireless platforms that can adapt not only their signals but also the physical locations from which they radiate and observe. This article presents pinching-antenna systems (PASS) as a spatially adaptive platform for ISAC. PASS adopts dielectric waveguides as signal-transport media and reconfigurable dielectric pinching antennas to realize programmable radiation and observation points along these waveguides. This unique architecture introduces new spatial degrees of freedom, allowing the network to dynamically reconfigure its interaction geometry with users, targets, and the surrounding environment, rather than solely optimizing signals over a fixed antenna geometry. Leveraging a terahertz testbed, we experimentally validate the spatial control and spatially selective reception capabilities of PASS and demonstrate constructive field combining between two coherent radiation points. We further explore how such spatial reconfigurability can enable radio-map-assisted communication, environment division multiple access, and closed-loop active radio perception. Finally, we identify key challenges and research directions toward practical PASS-ISAC networks.
\end{abstract}


\section{Introduction}
Future sixth-generation (6G) networks are expected to move beyond traditional information-delivery infrastructures toward intelligent wireless systems that can perceive and interact with the physical world \cite{nokia2024envisioning6gfuture}. Emerging applications such as digital twins, embodied artificial intelligence (AI), autonomous mobility, collaborative robotics, and immersive extended reality require wireless networks not only to provide reliable connectivity, but also to support accurate localization, sensing, and environmental awareness \cite{karapantelakis2024cyberphysical}. These emerging requirements are driving the convergence of communication, localization, sensing, and physical-world perception into future integrated wireless systems. In this context, integrated sensing and communication (ISAC) has emerged as a key enabler for seamlessly bridging communication networks and the physical world by jointly supporting efficient information transfer and radio-based perception within a unified framework \cite{yuan2025isacserieseditorial}.

Existing ISAC research has established a rich foundation spanning system architecture, waveform design, channel estimation, joint beamforming and localization, and resource optimization, demonstrating the feasibility and value of combining communication with radio-based perception. However, these advances still leave open a more fundamental question: \textit{Can a spatially reconfigurable platform be developed to provide a physical foundation for ISAC-oriented signal processing and resource allocation?}

To realize such a promising vision of ISAC-enabled interaction with the physical world, a suitable spatial platform candidate should satisfy four essential requirements. First, it must provide spatial awareness, namely, the ability to infer information about users, objects, and the surrounding environment from propagating wireless signals. 
Second, it must provide spatial control, enabling the system to proactively adapt the effective wireless channel by reconfiguring its radiation and observation geometry in response to acquired spatial information.
Third, these two capabilities must be tightly linked through real-time coordination or closed-loop adaptation, allowing perception to directly guide transmission and sensing behaviour. Finally, the platform must be practically deployable in future 6G networks, offering low-latency reconfiguration, high reliability, scalability, and autonomous operation.

Several existing spatial wireless technologies address different facets of this problem. Specifically, movable-antenna (MA) and fluid-antenna system (FAS) architectures introduce additional spatial degrees of freedom (DoF) by reconfiguring antenna positions or switching among candidate antenna ports, which improves both sensing and localization performance by adaptively exploiting spatial channel variations \cite{zhu2024movableantennas,new2025fluidantennas}. However, their reconfiguration range is typically limited to a few wavelengths, which constrains their ability to adapt across broader spatial regions and under more practical heterogeneous propagation geometries.
Furthermore, reconfigurable intelligent surfaces (RISs) primarily alter the propagation environment by tuning their reflection coefficients and amplitudes to establish virtual end-to-end line-of-sight (LoS) links, making them highly effective for efficient environment shaping \cite{hu2021robustSecureSumRateIRS}. However, their propagation modification ability depends strongly on the RIS deployment location and surrounding geometry, which limits their inherent ``spatial awareness" and makes sensing-driven adaptation indirect. In contrast,
distributed antenna systems (DASs) improve coverage by placing radiating nodes at geographically distributed locations \cite{zhang2020das5g}. Their macro-diversity can also support high-accuracy localization and environmental sensing through coordinated multi-point measurements. However, individual antenna positions generally remain fixed after deployment, limiting the runtime adaptability of their physical geometry to time-varying propagation conditions. Moreover,
leaky-wave antennas (LWAs) exploit guided-wave radiation along a waveguide structure, enabling controllable beam scanning through frequency variation or structural reconfiguration \cite{javanbakht2021reconfigurableLWA}. However, their spatial flexibility is still largely constrained by the underlying leakage structure and operating frequency, limiting their reconfigurable capabilities. 
Taken together, these technologies each address complementary but only part of the broader spatial-adaptation problem.  
This motivates pinching-antenna (PA) systems (PASS) as a candidate platform that can more naturally unify spatial awareness and spatial control within a single reconfigurable physical architecture suitable for future 6G systems \cite{fukuda2022pinching}.

Unlike prior PASS studies that primarily focus on communication-oriented feasibility~\cite{li2026unlocking}, modeling ~\cite{liu2025pinchingarchitecture}, or optimization~\cite{blockage}, we experimentally validate two key capabilities of PASS-enabled ISAC: spatial control and spatially selective
reception, exploiting a terahertz (THz) testbed. We further demonstrate the potential of constructive combining between two PAs, which underpins pinching beamforming. These experiments provide proof-of-concept evidence that PASS can dynamically reconfigure both its radiation and observation geometries, thereby motivating its potential applications in communication, localization, and sensing.


\section{PASS as a Spatially Adaptive Platform}

\subsection{From PAs to Programmable Radiation Points}
The concept of PASS was first demonstrated by NTT DOCOMO as a practical approach to radiate wireless signals from a dielectric waveguide to create communication zones at desired locations \cite{fukuda2022pinching}. In particular, a dielectric waveguide serves as the signal transmission medium, while a small separate dielectric element, typically mounted on a clip or clothespin, cf. Fig. \ref{fig:PASS_framework}, serves as the PA. Specifically, when the PA is brought into close proximity to the waveguide, electromagnetic energy couples from the guided mode into the PA and is radiated into free space. When the PA is released, the radiation ceases immediately. This simple physical mechanism makes it possible to activate or deactivate radiation locally along the waveguide and to rapidly reconfigure the coverage region in response to the actual propagation environment and user locations \cite{fukuda2022pinching}.
More recently, \cite{li2026unlocking} showed that the geometry and orientation of a PA can dynamically control its radiation direction. However, these studies have not directly shown how PA relocation modifies the spatial distribution of the transmitted and received signals. This behavior is important for PASS-enabled ISAC as it determines where electromagnetic energy can be delivered and where it can obtain spatially informative observations.

From a system perspective, the key innovation of PASS lies not merely that a dielectric waveguide can radiate, but that its effective radiation and potential observation points can be programmed over an extended spatial region. This capability shifts the design paradigm from optimizing beam patterns generated by fixed radiating structures to configuring the spatial locations where the wireless platform radiates and collects signals, thereby introducing a fundamentally new spatial DoF for controlling the wireless interaction geometry.

\subsection{From Programmable Radiation Points to Spatial Intelligence}
\begin{figure}
    \centering
    \includegraphics[width=\linewidth]{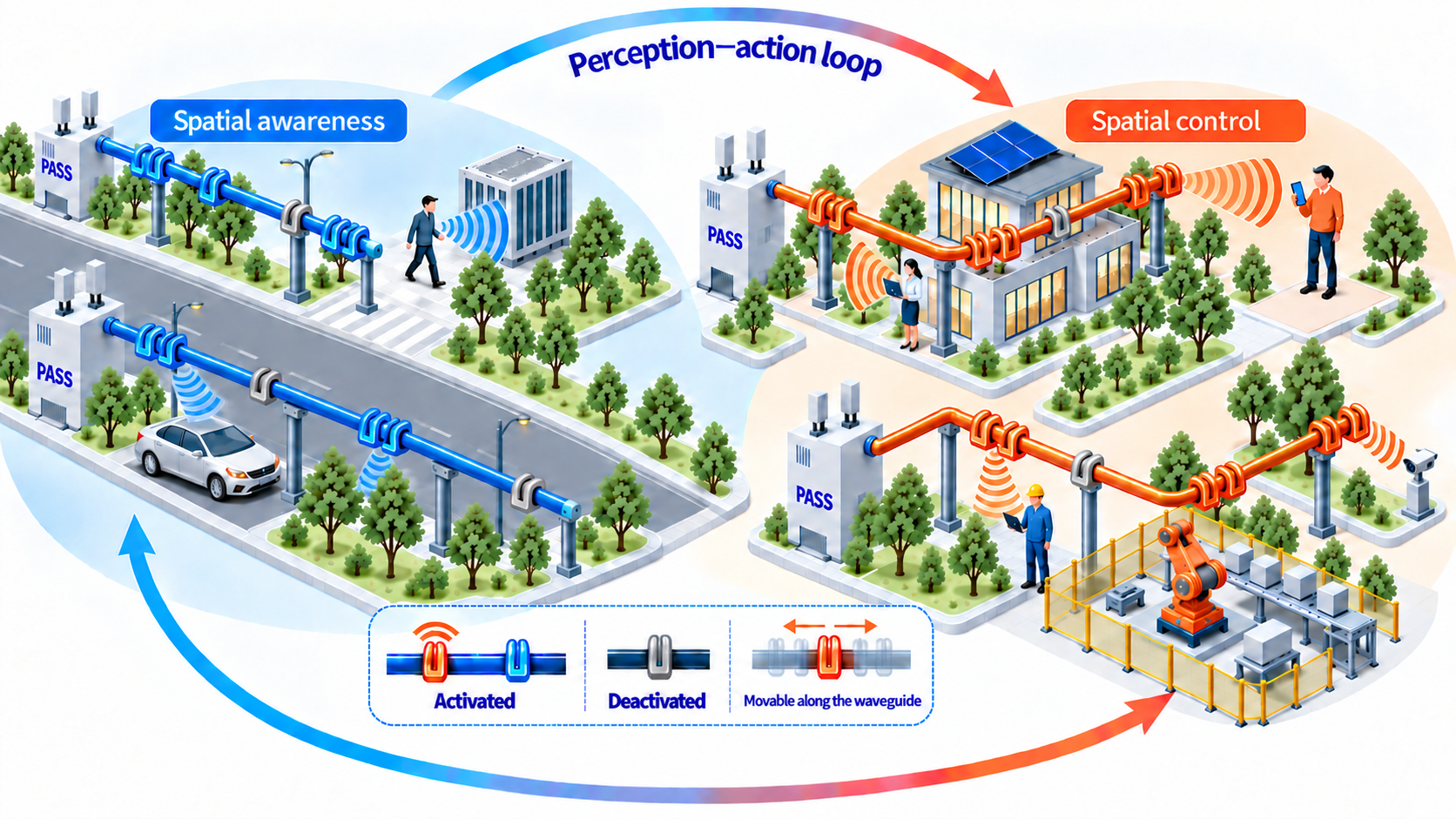}
\caption{PASS as a spatially adaptive platform for ISAC.}
    \label{fig:PASS_framework}
\end{figure}
The new spatial DoF introduced by PASS enables wireless systems not only to communicate with the physical world but also to actively observe and shape their interactions with it. As shown in Fig. \ref{fig:PASS_framework}, this capability arises since the programmable radiation points can naturally serve two complementary roles: programmable viewpoints for spatial awareness and programmable radiators for spatial control.

\textbf{Programmable viewpoints for spatial awareness:}
Spatial awareness relies on acquiring spatially informative observations of users, targets, and the surrounding environment. Such observations depend not only on signal processing algorithms, but also on the corresponding spatial geometry from which wireless measurements are acquired. PASS provides this capability by transforming programmable radiation points into programmable observation viewpoints, allowing the wireless platform to actively determine where spatial observations are made.
This flexibility is particularly important for effective sensing. In particular, nearby viewpoints can provide stronger echoes and enhanced spatial sensitivity for accurate target probing, whereas spatially distributed viewpoints offer wider coverage, richer angular diversity, and improved robustness against potential blockage and occlusion~\cite{blockage}. However, conventional wireless systems acquire measurements from fixed antenna locations, rendering their observation geometry largely predetermined after deployment. In contrast, PASS renders the observation geometry itself programmable. By activating different PAs along the waveguide, the wireless platform can proactively adapt its sensing viewpoints to support localization, channel sensing, target detection, and environmental perception.

\textbf{Programmable radiators for spatial control:}
Spatial control determines how the wireless platform actively shapes its electromagnetic interaction with the physical environment. Specifically, PASS provides this capability through programmable radiation points, which allow both the radiation locations and associated radiation characteristics to be dynamically configured. At a coarse spatial scale, activating or relocating PAs determines where the system radiates, thereby reconfiguring its coverage, link geometry, and target-illumination regions according to communication or sensing objectives. At a finer spatial scale, the relative positions of multiple active PAs jointly determine their free-space propagation distances, guided-wave phases, and relative radiation phases. These spatial and phase responses jointly constitute the additional DoF available for pinching beamforming, enabling more precise shaping of the resulting electromagnetic fields. Existing wireless systems, e.g., \cite{hu2021robustSecureSumRateIRS,zhu2024movableantennas,new2025fluidantennas}, primarily optimize electromagnetic fields generated by fixed radiators. PASS goes one step further by dynamically reconfiguring the radiation geometry itself, extending spatial control from beam adaptation over a fixed geometry to joint adaptation of the radiator locations and the resulting field distribution.                    

\textbf{From perception to action:}
Spatial awareness and spatial control are not independent capabilities; rather, they naturally complement each other. In particular, spatial awareness provides the information required to understand users, targets, and the surrounding environment, while spatial control determines how the wireless platform should respond to this information through programmable radiation points. As a result, environmental perception could directly guide subsequent wireless transmission and sensing tasks, transforming communication and sensing from largely open-loop operation into adaptive interactions with the physical world.
Indeed, existing wireless systems typically perform sensing and communication over a fixed physical geometry, making perception and transmission only loosely coupled. In contrast, PASS strengthens this coupling by allowing both observation locations and radiation locations to be reconfigured at runtime. The joint availability of spatial awareness and spatial control therefore elevates PASS from a reconfigurable communication platform to a proactive perception-action platform, providing the physical foundation for future closed-loop ISAC systems.

\section{Experimental Validation}
\begin{figure}[t]
\centerline{\includegraphics[width=\columnwidth]{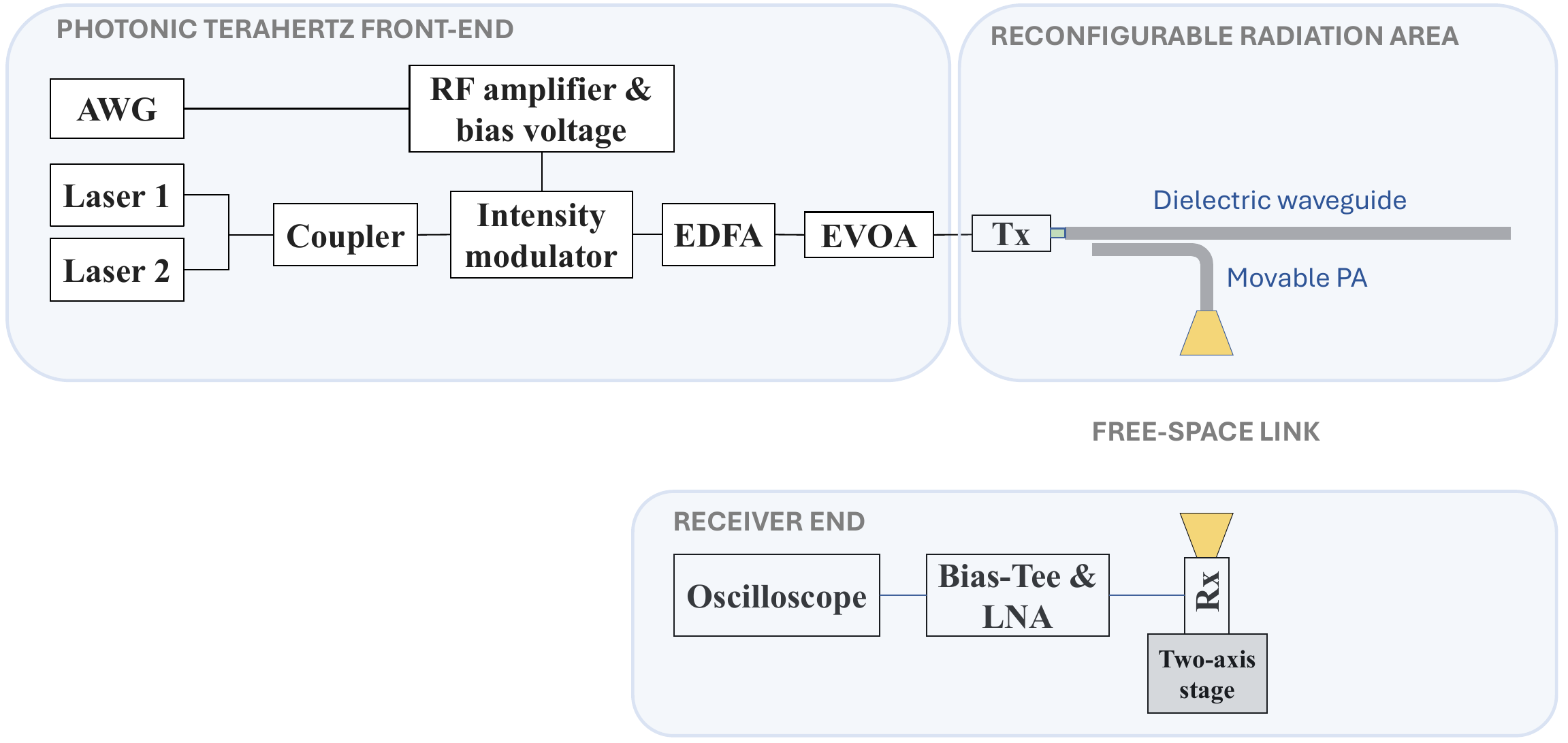}}
\vspace{-1em}
\caption{Schematic of the photonics-based terahertz system (EDFA: erbium-doped fiber amplifier, EVOA: electronic variable optical attenuator, Tx: transmitter, Rx: receiver, LNA: low-noise amplifier) \cite{wang2024demonstration}.}
\vspace{-1.5em}
\label{fig: ExperSys1}
\end{figure}
Although PASS was demonstrated in the millimeter-wave (mmWave) band \cite{fukuda2022pinching,li2026unlocking}, its underlying architecture is not intrinsically restricted to a particular frequency band. The THz band is particularly attractive for future 6G systems because its abundant bandwidth can support extremely high data rates, while its short wavelength enables compact radiating structures, high spatial resolution, and highly directional propagation. These properties are also beneficial for ISAC, where fine-grained control of the wireless interaction geometry and accurate spatial observations are essential. Motivated by these advantages, we experimentally investigate whether PASS can realize its intended spatial capabilities in the THz band.

Specifically, we conduct three experiments using the THz testbed shown in Fig.~\ref{fig: ExperSys1} and the experimental configurations illustrated in Figs.~\ref{fig:experimental_setup} and~\ref{fig:constructive_setup}. Specifically, Fig.~\ref{fig: Erx} demonstrates spatially selective reception, Fig.~\ref{fig: Etx} demonstrates spatial control, and Fig.~\ref{fig:constructive_map} demonstrates constructive combining between two coherent radiation points. All three experiments used the same measurement settings to ensure a consistent basis for comparison, and the measured responses are presented as signal-to-noise ratios (SNRs) relative to the measured noise level. In these figures, black circles indicate the PA locations, whereas white circles mark the locations of the maximum measured SNR.

\subsection{Photonics-Based Terahertz Testbed}
The measurement setup where the PA acts as a wireless transmitter is shown in Fig.~\ref{fig: ExperSys1}. Two tunable lasers operating near $1,550$ nm are first combined and modulated by an electrical waveform generated using an arbitrary waveform generator (AWG). At the transmitter, the modulated optical carriers are photomixed in a uni-traveling-carrier photodiode (UTC-PD) to generate a $125$-GHz electrical carrier with an output power of $-3$~dBm. At the receiver, a $25$-dBi horn antenna is connected to a D-band zero-bias Schottky diode for direct detection of the received terahertz signal. The detected baseband signal is captured by an oscilloscope. The PA can also operate as a receiver by interchanging the  Tx and Rx configurations in Fig. ~\ref{fig: ExperSys1}. The carrier frequency, modulation waveform, transmitter settings, and receiver gain were kept unchanged across all experiments to ensure consistent measurement conditions.

\begin{figure}[!t]
    \centering
    \subfloat[PA operating as the transmitter.]{
        \includegraphics[width=0.49\columnwidth]{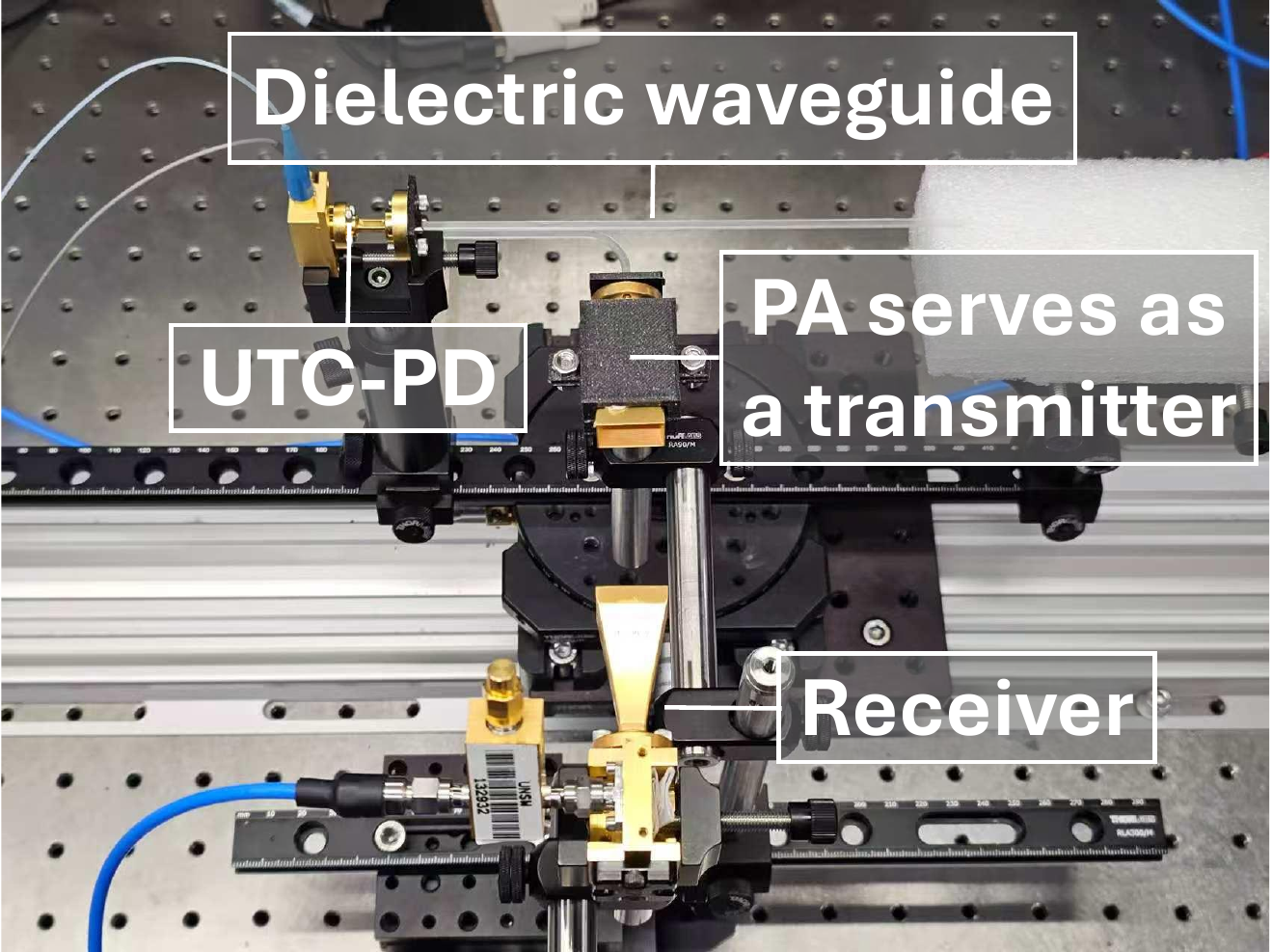}
        \label{fig:pa_tx}
    } 
    \subfloat[PA operating as the receiver.]{
        \includegraphics[width=0.49\columnwidth]{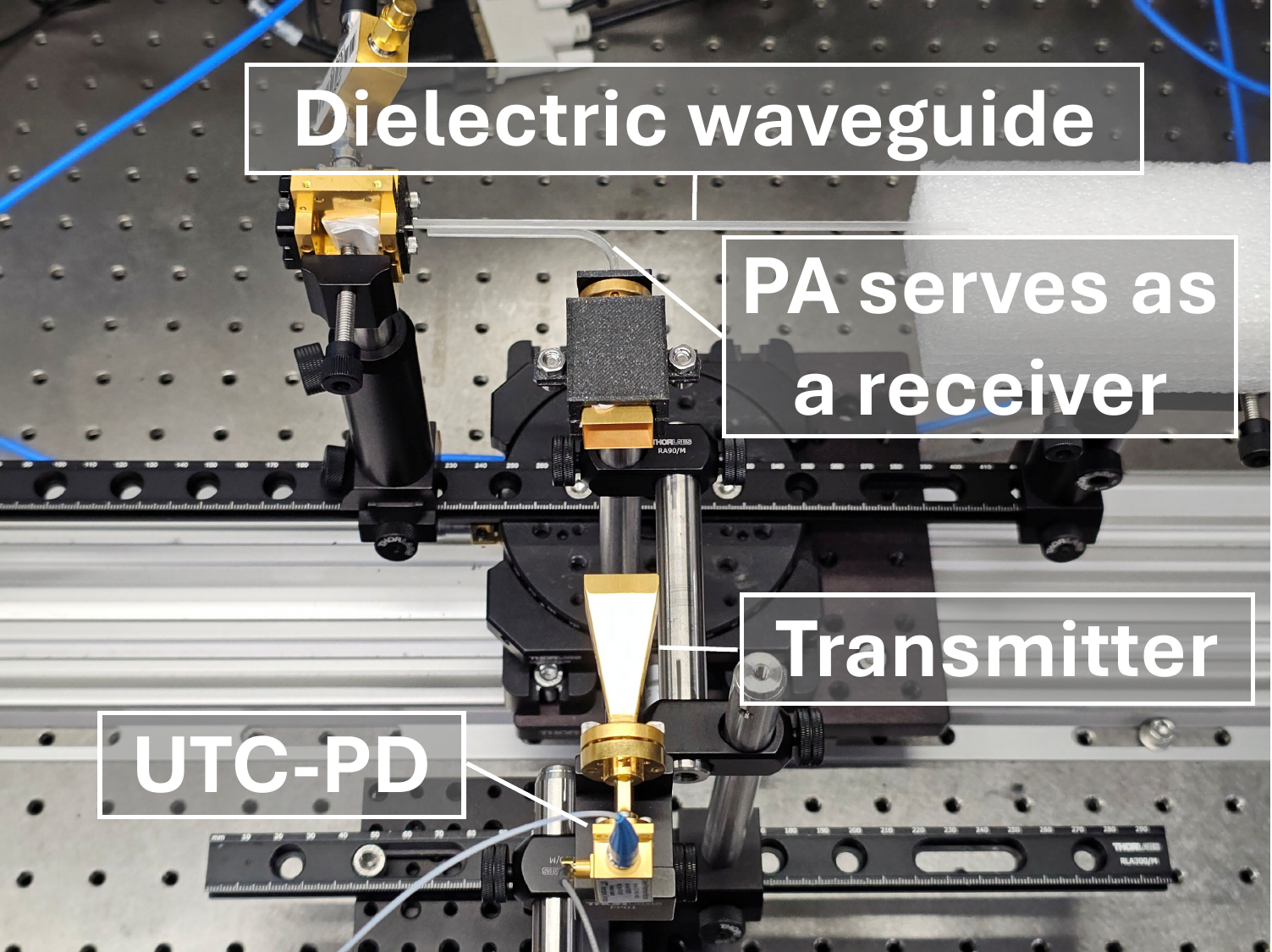}
        \label{fig:pa_rx}
    }
    \caption{Experimental setup for validating the spatial control of the proposed PASS: (a) the PA operates as the transmitter; (b) the PA operates as the receiver.}
    \label{fig:experimental_setup}
\end{figure}

The dielectric waveguide and the PA were fabricated by three-dimensional printing with cyclic olefin copolymer (COC) filament, with a rectangular cross section of $1.90$~mm $\times 1.02$~mm. The PA had an L-shaped configuration, with one section positioned alongside the waveguide over a $50$~mm-long coupling region. The coupling gap between the PA and the waveguide was kept below $1$~mm to ensure efficient electromagnetic coupling, while the other section was aligned with the horn antenna. The PA was configured as the transmitter in Fig.~\ref{fig:pa_tx} and as the receiver in Fig.~\ref{fig:pa_rx}.

\subsection{Validation of Spatially Selective Reception and Control}
The waveguide feed point was located at $(0,3)$~mm in the measurement coordinate system. In both the transmit and receive experiments, the PA was fixed at two positions, $(10,3)$~mm and $(40,3)$~mm. The scan was performed in the vertical $x$-$y$ plane, with the $y$-axis normal to the horizontal plane. When the PA operated as the transmitter, the receiving horn was scanned over the measurement region. Conversely, when the PA operated as the receiver, the transmitting horn was scanned over the same region to characterize the spatially-dependent reception response. The scan adopted step sizes of $2$~mm and $1$~mm along the $x$- and $y$-axes, respectively.

\begin{figure}[t]
\centerline{\includegraphics[width=\columnwidth]{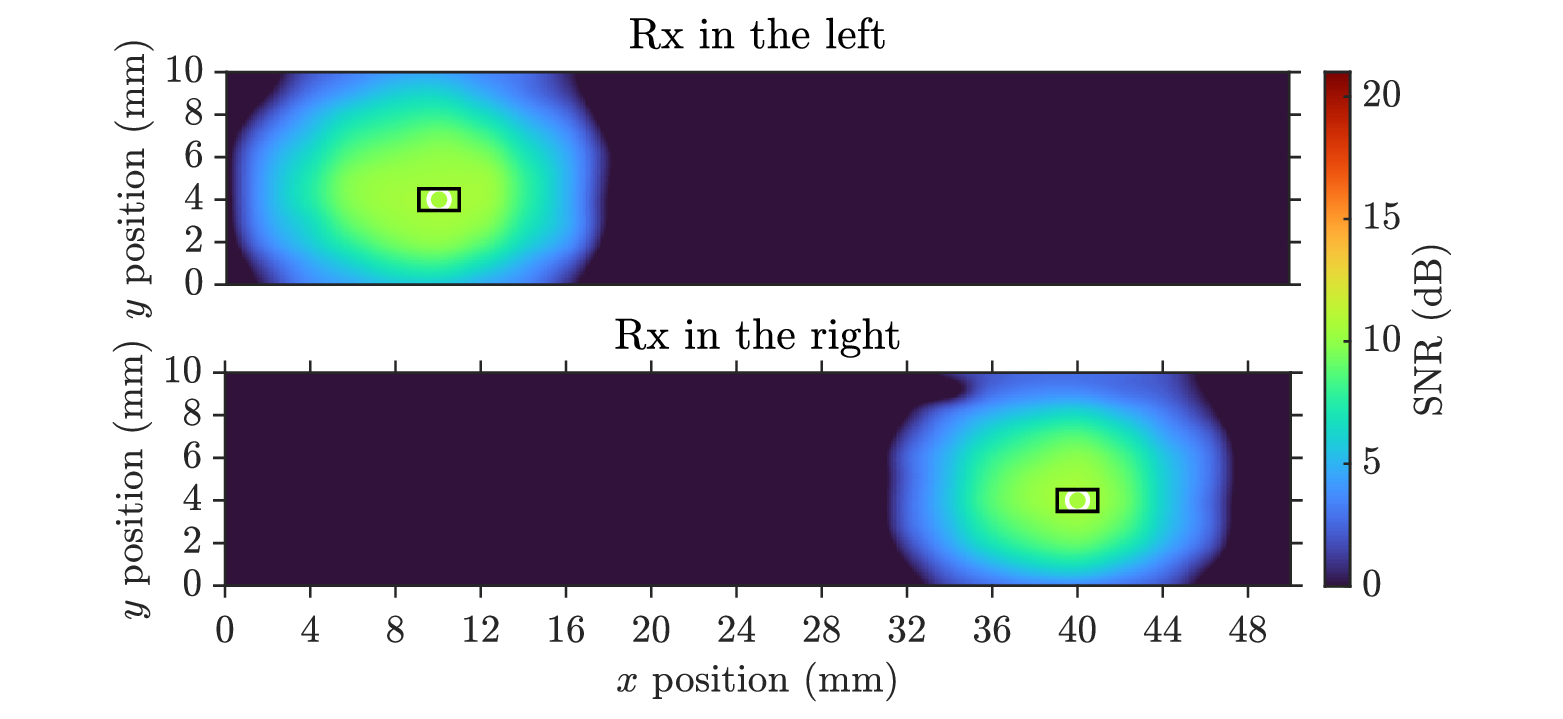}}
\vspace{-1em}
\caption{Measured SNR maps for a receiving PA positioned on the left (top) and right (bottom).}
\vspace{-1em}
\label{fig: Erx}
\end{figure}

\begin{figure}[t]
\centerline{\includegraphics[width=\columnwidth]{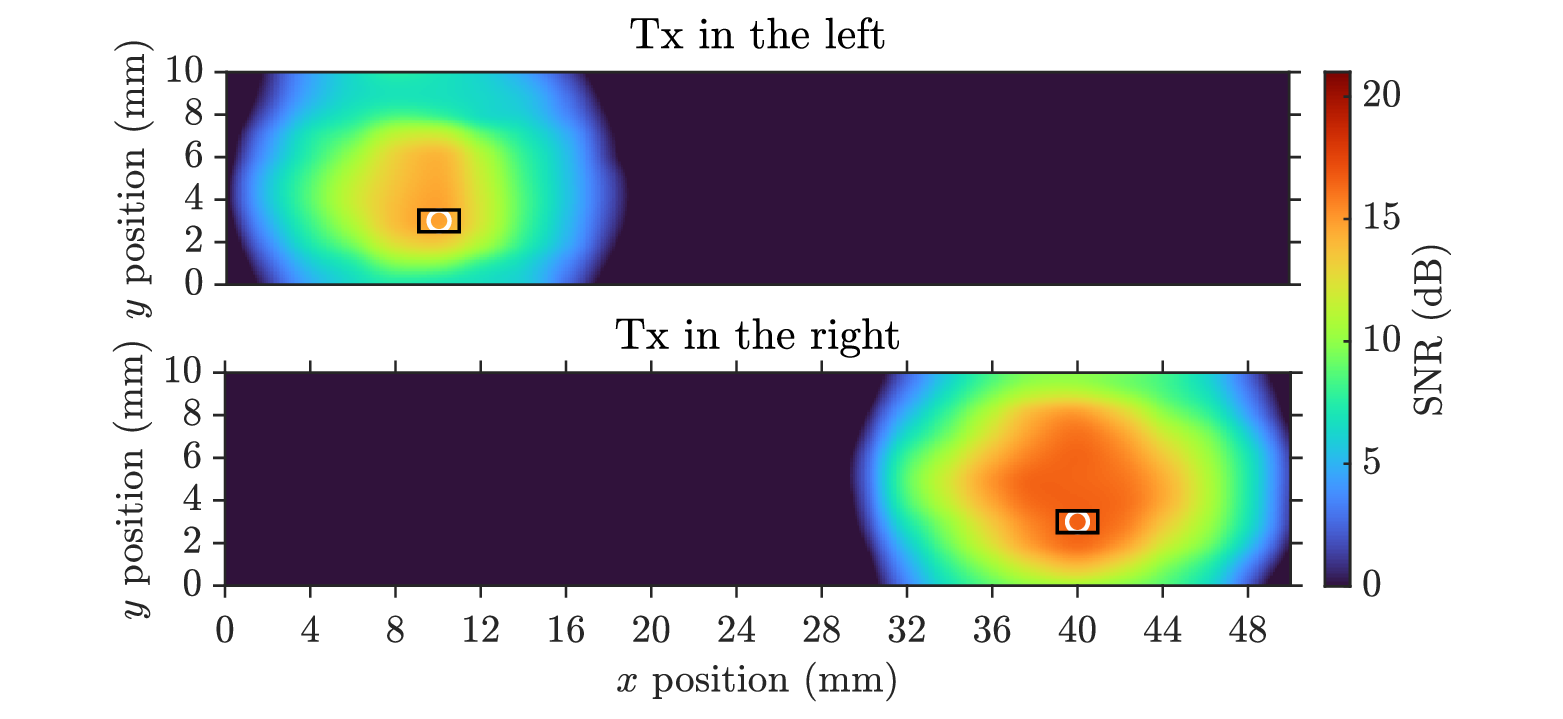}}
\vspace{-1em}
\caption{Measured SNR maps for a transmitting PA positioned on the left (top) and right (bottom).}
\vspace{-1em}
\label{fig: Etx}
\end{figure}

\subsubsection{Spatially Selective Reception}
Fig.~\ref{fig: Erx} demonstrates that the PA can serve as a reconfigurable observation point. Displacing the PA by $30$~mm shifts the region of strongest reception by the same distance. Moreover, comparing the two SNR maps shows that, at a fixed transmitting-horn location, the received SNR can differ by more than $11$~dB depending on whether the PA is positioned near that location or moved to the other position. Consequently, each PA position acts as a spatially selective observation channel, and the power measured at a common source location becomes strongly dependent on the observation geometry rather than on the signal processing alone. This constitutes the physical basis for spatial awareness, since a target or clutter illuminating the platform is observed with markedly different strengths from different PA locations, thereby offering valuable spatial diversity that can be exploited for creating effective localization, target discrimination, and environmental sensing.

\subsubsection{Spatial Control}
As shown in Fig.~\ref{fig: Etx}, relocating the transmitting PA from $x=10$~mm to $x=40$~mm shifts the high-SNR region by $30$~mm along the $x$-axis, while leaving its peak position along the $y$-axis unchanged. The displacement of the high-SNR region therefore closely coincides with the PA movement. After the PA is moved, the received signal falls to the noise floor at the original PA location, resulting in a spatial contrast exceeding $15$~dB. These results confirm that the effective radiation point can be reconfigured simply by altering the PA position.
Moreover, despite having the same coupling length and gap, the PA located farther from the waveguide feed point exhibits a stronger peak response. This difference indicates that the radiated signal strength does not decrease monotonically with distance from the feed point. Instead, it depends on the local guided-wave field at the PA position, including its amplitude and phase.
Consequently, a PA located closer to the feed point may still radiate more weakly if positioned at a location with an unfavourable guided-wave field. Hence, the PA position affects not only the location of the radiation point but also the effective channel gain, providing an additional spatial DoF for controlling the radiated power in PASS. 

\subsection{Validation of Constructive Field Combining}
\begin{figure}[t!]
    \centering
    \subfloat[Experimental setup.]{
        \includegraphics[width=0.7\linewidth]{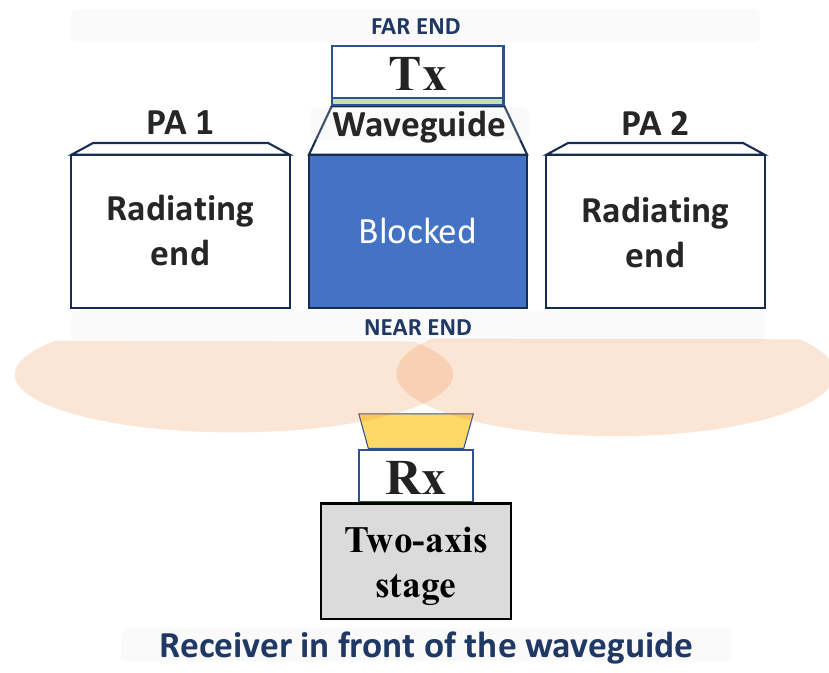}
        \label{fig:constructive_setup}
    }
    \hfill \hspace{-3mm}
    \subfloat[Measured SNR map for constructive combining using two PAs.]{
        \includegraphics[width=0.7\linewidth]{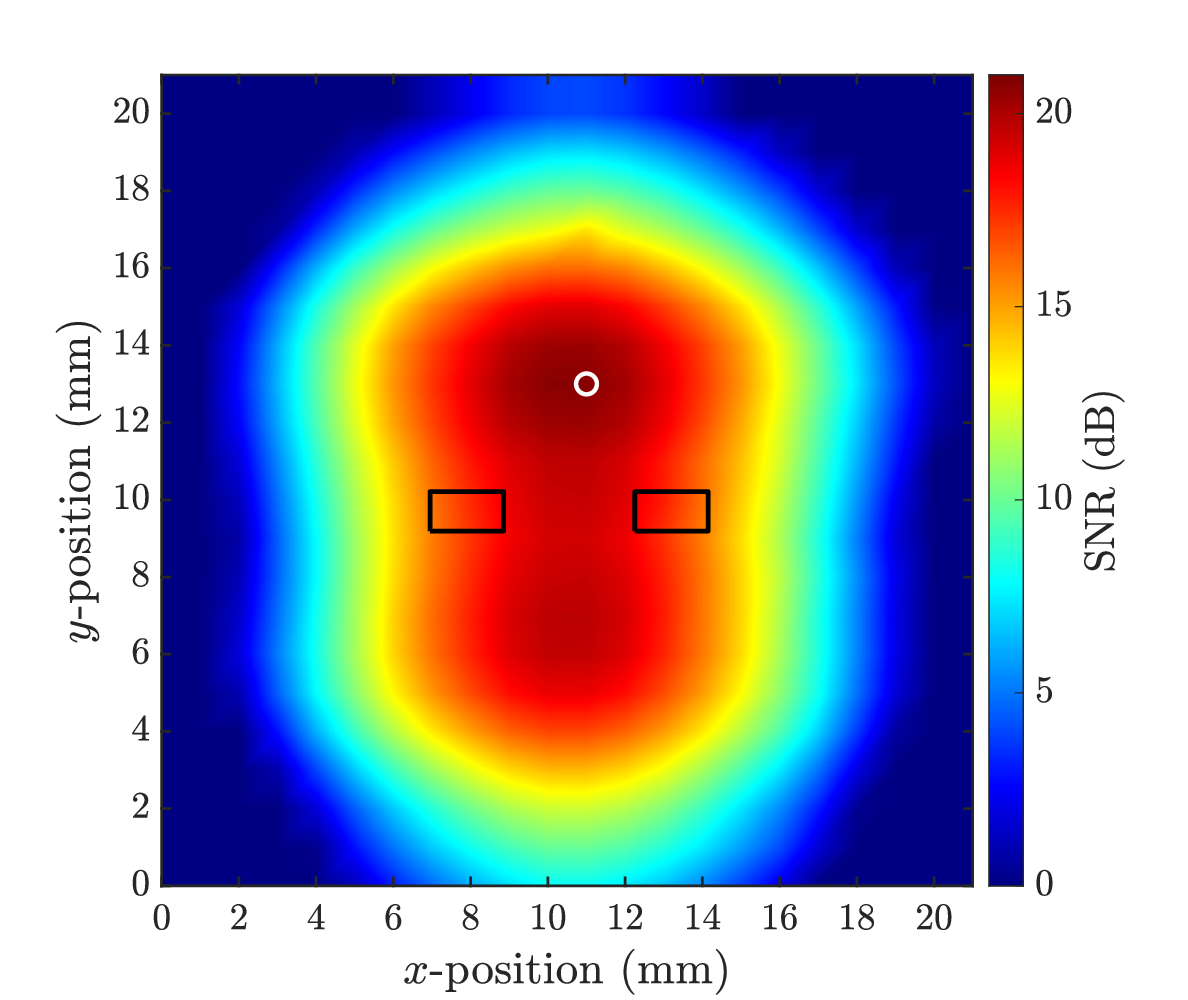}
        \label{fig:constructive_map}
    }
    \caption{Experimental validation of constructive combining using two PAs.}
    \label{fig:constructive_combining}
\end{figure}
To validate constructive combining, the experimental configuration is shown in Fig.~\ref{fig:constructive_setup}, which provides a front view of the setup. Two PAs are positioned on opposite sides of the waveguide. 
The array is excited through its middle core, while the output of this core is blocked to suppress direct radiation. 
The coupling gaps between each side core and the middle core are $0.68$ mm. The resulting field is measured by a receiving horn positioned $20$~mm in front of the two radiating ends, exploiting the same transceiver chain and transmit power adopted in the preceding experiments.

 Fig.~\ref{fig:constructive_map} demonstrates constructive combining of two radiation points. For a single PA, the strongest received power occurs near the PA position, as shown in Fig.~\ref{fig: Etx}. When the two PAs are activated simultaneously, the peak response is formed between the two radiation points and is higher than that obtained with a single PA. This indicates that the two radiated fields add constructively in the intermediate region. This controllability provides the physical basis for future pinching beamforming~\cite{blockage}, in which the PA positions and radiated powers can be jointly optimized to focus energy toward a desired location.



\section{Implications for Future Networks}
PASS-ISAC opens new opportunities for future wireless networks by jointly exploiting spatial reconfigurability and environmental perception. This section discusses several emerging network paradigms and system-level functionalities enabled by this integration.
\subsection{Radio Map}
A radio map is a site-specific database of slowly varying, location-dependent channel characteristics, such as the average received signal strength, path loss, and LoS availability. By retrieving such information based on location, rather than repeatedly estimating it through resource-hungry pilot signalling, a radio map can reduce the online training overhead at the expense of offline map construction and storage. Within the ISAC paradigm, the sensing functionality can be leveraged to construct and update the map, whereas the communication functionality can exploit it for proactive beam management and blockage-aware transmission. Meanwhile, flexible-antenna systems further extend the dimensionality of the map by introducing antenna repositioning as an additional spatial DoF, thereby unlocking spatial reconfiguration that benefits both functionalities~\cite{shao2025tutorial}. Indeed, PASS exemplifies this concept: relocating a PA along the waveguide jointly alters the in-waveguide propagation response and the free-space channel toward the served users and targets. By identifying antenna positions that provide favorable communication links and target illumination, a placement-aware radio map can guide effective antenna relocation and enable joint optimization of communication and sensing performance. Efficient low-cost construction and timely updates, however, remain open research problems.

\subsection{Environment Division Multiple Access (EDMA)}
By rendering the wireless channel dynamically reconfigurable rather than static, flexible-antenna technologies open new opportunities for environment-aware multiple access. A prominent example unlocked by PASS is EDMA, which separates users in the environment domain instead of the conventional time, frequency, or code domains, by exploiting propagation characteristics that are pronounced at mmWave and THz frequencies, such as LoS blockage and severe large-scale path loss~\cite{ding2026environment}. Through judicious positioning of the PAs, a favorable LoS link is established to the intended users, whereas the links toward the remaining users are obstructed or severely attenuated. Consequently, inter-user interference is largely suppressed by the propagation environment itself, allowing multiple users to reuse the same time-frequency resources with reduced reliance on orthogonal resource partitioning. The same controllability of blockage and propagation geometry is also appealing for physical-layer security provisioning, since confining the LoS link to the legitimate user while denying a favorable path to potential eavesdroppers safeguards confidentiality with reduced reliance on artificial noise and transmit power~\cite{blockage}.
Realizing these gains, however, requires accurate knowledge of the surrounding environment, which is precisely what the sensing functionality of ISAC offers. By characterizing the prevailing blockage and large-scale path loss, sensing informs the placement of the PAs to achieve the desired user separation, thereby positioning EDMA as a compelling application for PASS-enabled ISAC networks.

\subsection{Closed-Loop Active Radio Perception}
Closed-loop active radio perception is expected to become increasingly important in 6G networks, where communication services will be deeply coupled with dynamic physical processes, such as autonomous mobility, collaborative robotics, and industrial automation. In these scenarios, sensing is not an end in itself: the acquired spatial information must continuously guide how and where the network interacts with users, targets, and the surrounding environment. This motivates a closed-loop architecture in which radio observations are translated into physical-layer actions, whose outcomes are subsequently perceived and used for further adaptation.
PASS-ISAC provides a natural physical platform for realizing this vision. Its programmable PAs can serve as spatially distributed viewpoints to acquire user, target, and environmental information, while also acting as reconfigurable radiators that modify the communication and sensing geometry. For example, the network may first use distributed PAs to localize a moving user or detect an emerging blockage, and then activate a more favorable PA or relocate an existing PA to maintain a strong LoS link. Similarly, target information obtained from one set of viewpoints can guide the selection of new PA locations for stronger illumination or more informative observations. The resulting measurements can again update the network's spatial knowledge, naturally forming a perception–decision–action cycle.
Such a capability moves ISAC beyond simply sharing communication and sensing resources toward continuous interaction with the physical world. By allowing perception to directly determine where the network should observe and radiate next, PASS-ISAC makes spatial geometry an adaptive network resource, opening a viable path toward more autonomous and environment-aware 6G systems.

\section{Conclusions, Open Challenges, and Future Directions}
PASS provides a physical architecture in which both observation and radiation points are programmable, thereby introducing spatial awareness and spatial control as complementary capabilities. Using a conventional photonics-based THz testbed, we validate these capabilities experimentally: relocating a receiving PA shifts the region of strongest reception and enables spatially selective observation; relocating a transmitting PA shifts the high-SNR radiation region; and two coherently excited PAs produce constructive field combining. Together, these results show that PASS can turn spatial geometry from a fixed hardware constraint into an adaptive network resource, allowing the system to dynamically reconfigure where it observes and radiates. Realizing this potential in practical PASS-ISAC networks, however, requires advances from physical hardware to system-level design. Key open challenges and research directions are outlined below:

\subsection{Sensing-Aware Electromagnetic Modeling and Calibration}
Most existing PASS studies adopt simplified system-level models in which a PA is characterized primarily by its position-dependent propagation distance, phase, and radiated power. Although such models are useful for capturing the basic spatial DoF of PASS, they are generally insufficient for PASS-ISAC. 
In practical communication-only PASS, variations caused by the propagation environment and hardware state can often be captured in the effective communication channel without the need to identify their physical origins.
However, in PASS-ISAC, distinguishing these origins becomes essential as the received signal is also exploited to infer the physical environments.
For example, an incorrectly positioned PA changes the effective transmit or receive location, producing phase, delay, and spatial-response variations that may be misinterpreted as target motion. Likewise, variations in waveguide coupling or waveguide loss reduce the received echo power, which may be mistaken for blockage or increased propagation loss.
If these effects are not properly separated, communication algorithms may operate with inaccurate effective CSI, while sensing algorithms may misinterpret hardware variations as changes in target position, range, angle, or blockage state. Therefore, an important open challenge is to develop practical PASS-ISAC models that capture both hardware and environmental variations while enabling their respective contributions to be separated. These models must also remain sufficiently simple for system-level performance analysis and practical design.
\subsection{Hardware Impairments and Robust PASS-ISAC}
Even with an accurately calibrated nominal model, a practical PASS may not consistently realize its intended electromagnetic response during operation. This discrepancy arises not only from hardware drift but also from the physical reconfiguration process. For example, the actual PA position may deviate from its commanded location, while variations in PA attachment, orientation, or waveguide conditions can alter the coupling response. Manufacturing tolerances can further cause nominally identical PAs to exhibit different responses, and repeated detachment and reattachment may not reproduce the calibrated amplitude and phase. These effects are particularly pronounced in high-frequency and near-field PASS deployments, where small positional errors or mechanical vibrations can induce significant phase variations. In multi-PA operation, such errors may accumulate and degrade coherent beamforming and focusing.
Because PASS-ISAC relies on a reconfigurable PA infrastructure for both communication and sensing, these hardware uncertainties should be incorporated into the system design rather than treated as negligible implementation errors. Future work should develop robust PASS-ISAC designs and practical calibration strategies that remain reliable under time-varying hardware conditions without incurring excessive overhead.

\subsection{Channel Estimation for PASS-ISAC}

Although the need to acquire channel state information (CSI) over a large number of possible antenna configurations is also encountered in other reconfigurable antenna-assisted systems, PASS introduces an additional complication: moving a PA changes both the free-space link to the user and the in-waveguide response from the feed to the radiation point~\cite{liu2025pinchingarchitecture}. With multiple PAs, the overall response also depends on their activation patterns, relative positions, and guided-wave phases. Consequently, the objective is not simply to estimate a finite set of independent channels, but to infer the effective response over a large configuration space from a limited number of measurements. This is challenging because the free-space geometry may vary smoothly with the PA position, whereas the guided-wave phase and amplitude can change substantially along the same waveguide. Developing accurate predictors for this coupled response without exhaustive pilot measurements remains an open problem.
PASS-ISAC further changes what constitutes an informative channel measurement. Communication-oriented acquisition may favor PA locations with strong effective channels, whereas sensing may benefit from distributed viewpoints and diverse propagation paths for target localization and environmental perception. Thus, the PA locations that are most useful for communication are not necessarily those that provide the most informative sensing observations. How to exploit these observations for channel prediction without requiring dense PA probing or perfect environmental knowledge remains an open challenge.
\subsection{Fast Pinching Deployment}
Despite the appealing spatial flexibility of PASS, its practical implementation involves a trade-off among positioning accuracy, response latency, and energy consumption. A PA can be repositioned along the waveguide to a favorable position that reduces propagation loss, improves phase alignment, and enhances channel quality. However, long-distance repositioning incurs considerable delay and energy consumption, which is particularly undesirable in user-tracking scenarios requiring real-time adaptation. Thus, whether a PA should move toward a more favorable position or remain near its current location becomes a critical design decision.
One alternative is to predeploy multiple candidate PAs along the waveguide and activate the one closest to the desired position. Although this approach reduces response latency, it discretizes the feasible PA positions, lowers positioning resolution, and reduces continuous placement to an antenna-selection problem, thereby sacrificing the spatial flexibility of PASS. Moreover, the energy costs of both PA movement and activation must be considered and modeled in an appropriate manner. Thus, a fast and low-complexity strategy is needed to strike an efficient balance among channel performance, repositioning latency, spatial resolution, energy consumption, and system complexity.

 \subsection{Resource Allocation}
PASS inherently gives rise to challenging resource allocation problems. Relocating a PA jointly changes the free-space propagation channel and the in-waveguide response. In particular, the guided-wave phase varies periodically with the PA position, while the PA-induced phase shift can also depend on the locally coupled amplitude. Consequently, the effective channel becomes a highly nonconvex function of the PA position, with multiple local optima rather than a simple monotonic distance-dependent response. The challenge becomes significantly more pronounced in PASS-ISAC-based resource allocation framework, in which generally unavoidable clutter reflections, the position-dependent multi-angle observation capability of the PAs, and their dynamic positioning along the waveguides render extended-target sensing a practically relevant objective. This enhanced capability, however, introduces new design challenges. For example, in Cramér–Rao bound (CRB)-driven designs, key sensing parameters, including viewing angles, phase gradients, and effective angular scattering weights, are jointly dictated by the PA positions and further coupled with the communication and sensing resources. Therefore, fully unlocking the potential of PASS-assisted ISAC hinges on the joint optimization of PA positioning, power-ratio allocation, waveguide beamforming, and sensing–communication resource allocation, while striking a balance between optimality and computational complexity, which remains an open research problem.

\bibliographystyle{IEEEtran}
\bibliography{reference}

@ARTICLE{blockage,
  author={Zhao, Ruotong and Hu, Shaokang and Mishra, Deepak and Ng, Derrick Wing Kwan},
  journal={IEEE Trans. Mob. Comput.}, 
  title={Robust and Secure Blockage-Aware Pinching Antenna-Assisted Wireless Communication}, 
  volume={25},
  number={10},
  pages={17547--17565},
  year = {Oct. 2026}
  }

@article{zhang2020das5g,
  author={Heath, Robert and Peters, Steven and Wang, Yi and Zhang, Jiayin},
  journal={IEEE Commun. Mag.}, 
  title={A current perspective on distributed antenna systems for the downlink of cellular systems}, 
  year={Apr. 2013},
  volume={51},
  number={4},
  pages={161-167},
  doi={10.1109/MCOM.2013.6495775}}

@article{javanbakht2021reconfigurableLWA,
  author  = {Javanbakht, N. and others},
  title   = {A review of reconfigurable leaky-wave antennas},
  journal = {IEEE Access},
  volume  = {9},
  pages   = {111742--111758},
  month   = {Aug.},
  year    = {2021}
}

@article{hu2021robustSecureSumRateIRS,
  author  = {Hu, Shaokang and Wei, Zhiqiang and Cai, Yuanxin and Liu, Chang and Ng, Derrick Wing Kwan and Yuan, Jinhong},
  title   = {Robust and Secure Sum-Rate Maximization for Multiuser {MISO} Downlink Systems with Self-Sustainable {IRS}},
  journal = {IEEE Trans. Commun.},
  volume  = {69},
  number  = {10},
  pages   = {7032--7049},
  month   = {Oct.},
  year    = {2021},
  doi     = {10.1109/TCOMM.2021.3097140}
}

@article{wang2024demonstration,
  author  = {Qigejian Wang and Yirui Deng and Deepak Mishra and Yixuan Xie and Elias Aboutanios and Shaghik Atakaramians},
  title   = {Demonstration of photonics-based {D}-band integrated localization and communication},
  journal = {Appl. Opt.},
  volume  = {63},
  number  = {15},
  pages   = {4068--4076},
  month   = may,
  year    = {2024},
  doi     = {10.1364/AO.516852}
}

@article{li2026unlocking,
  title={Unlocking Directional Radiation in Pinching-Antenna Systems: Geometry-Aware Design and Experimental Verification},
  author={Li, Haoyang and Liu, Weidong and Li, Zhongliang and Chen, Gaojie and Yang, Zheng and Ding, Zhiguo},
  journal={arXiv preprint arXiv:2607.24011},
  year={2026},
  archivePrefix={arXiv},
  eprint={2607.24011},
  primaryClass={eess.SY}
}

@article{liu2025pinchingarchitecture,
  author  = {Liu, Y. and others},
  title   = {Pinching-antenna systems: Architecture designs, opportunities, and outlook},
  journal = {IEEE Commun. Mag.},
  note    = {To be published},
  year    = {2025}
}

@article{new2025fluidantennas,
  author  = {New, W. K. and others},
  title   = {Fluid antenna systems: Redefining reconfigurable wireless communications},
  journal = {IEEE J. Sel. Areas Commun.},
  volume  = {43},
  number  = {3},
  pages   = {800--818},
  month   = {Mar.},
  year    = {2025}
}

@article{zhu2024movableantennas,
  author  = {Zhu, L. and others},
  title   = {Movable antennas for wireless communication: Opportunities and challenges},
  journal = {IEEE Commun. Mag.},
  volume  = {62},
  number  = {6},
  pages   = {114--121},
  month   = {Jun.},
  year    = {2024}
}

@misc{karapantelakis2024cyberphysical,
  author       = {{Ericsson}},
  title        = {{6G} -- Connectivity and Beyond for the 2030s},
  year         = {2024},
  howpublished = {White paper},
  url          = {https://www.ericsson.com/en/reports-and-papers/white-papers/6g-connectivity-and-beyond-for-the-2030s},
  note         = {Accessed: Sep. 17, 2026}
}

@misc{nokia2024envisioning6gfuture,
  author       = {{Nokia Bell Labs}},
  title        = {Envisioning a {6G} Future},
  year         = {2024},
  howpublished = {E-book},
  url          = {https://www.nokia.com/sites/default/files/2024-12/nokia_bell_labs_envisioning_a_6g_future_ebook_en.pdf},
  note         = {Accessed: Sep. 17, 2026}
}

@article{yuan2025isacserieseditorial,
  author       = {Yuan, Weijie and Alexandropoulos, George C. and Caire, Giuseppe and Cordeiro, Carlos and Gu, Tao and Wang, Pu},
  title        = {Series Editorial: Integrated Sensing and Communications},
  journal      = {IEEE Commun. Mag.},
  year         = {2025},
  month        = dec,
  pages        = {62--63}
}

@article{shao2025tutorial,
 author={Shao, Xiaodan and Mei, Weidong and You, Changsheng and Wu, Qingqing and Zheng, Beixiong and Wang, Cheng-Xiang and Li, Junling and Zhang, Rui and Schober, Robert and Zhu, Lipeng and Zhuang, Weihua and Shen, Xuemin},
  journal={IEEE Commun. Surv. Tutorials	},
  title={A Tutorial on Six-Dimensional Movable Antenna for {6G} Networks: Synergizing Positionable and Rotatable Antennas}, 
  year={2026},
  volume={28},
  number={},
  pages={3666-3709},
  doi={10.1109/COMST.2025.3602939}}

@article{ding2026environment,
  title={Environment division multiple access ({EDMA}): A feasibility study via pinching antennas},
  author={Ding, Zhiguo and Schober, Robert and Poor, H Vincent},
  journal={IEEE Trans. Wireless Commun.},
  volume={25},
  pages={15675-15691},
  year={Apr. 2026},
  publisher={IEEE}
}

@article{fukuda2022pinching,
  author  = {Atsushi Fukuda and Hiroto Yamamoto and Hiroshi Okazaki and Yasunori Suzuki and Kunihiro Kawai},
  title   = {Pinching Antenna---Using a Dielectric Waveguide as an Antenna},
  journal = {NTT DOCOMO Technical Journal},
  volume  = {23},
  number  = {3},
  pages    = {5--12},
  month    = jan,
  year     = {2022}
}

\section*{Biographies}
\textbf{Shaokang Hu} (Member, IEEE) is an Associate Lecturer, an Early Career Academic Fellow (ECAF), and a Postdoctoral Research Associate with the School of Electrical Engineering and Telecommunications (EE\&T), the University of New South Wales (UNSW), Sydney, Australia. She received the B.Eng. degree (with first-class Honours) from UNSW in 2017 and the M.E. and Ph.D. degrees in electrical engineering from UNSW in 2018 and 2024, respectively.

\textbf{Ruotong Zhao} (Graduate Student Member, IEEE) received the B.E. degree in telecommunications and the M.Phil. degree in electrical engineering from UNSW, in 2022 and 2025, respectively, where she is currently pursuing the Ph.D. degree. 

\textbf{Qigejian Wang} (Member, IEEE) is an Associate Lecturer, ECAF, and Postdoctoral Research Associate with the Terahertz Innovation Group in the School of EE\&T at UNSW. He received the B.Eng. degree from Huazhong University of Science and Technology (HUST) in 2017, followed by the M.Phil. and Ph.D. degrees in electrical engineering from UNSW in 2021 and 2025, respectively. 

\textbf{Shaghik Atakaramians} (Senior Member, IEEE; Senior Member, OPTICA) is an Associate Professor in the School of EE\&T at UNSW, where she leads the Terahertz Innovation Group. She received her Ph.D. in electrical and electronic engineering in 2011, earning the University Doctoral Research Medal and the Gertrude Rohan Memorial Prize.  
In 2024, she received the prestigious IRMMW-THz Zhenyi Wang Award for Excellence in recognition of her outstanding contributions to infrared, millimeter, and terahertz science. She also serves as an Associate Editor for APL Photonics.

\textbf{Derrick Wing Kwan Ng} (S’06-M’12-SM’17-F’21) is currently a Scientia Associate Professor with UNSW. He received his bachelor’s degree (with first-class Honours) and Master of Philosophy degree in electronic engineering from The Hong Kong University of Science and Technology (HKUST), Hong Kong, in 2006 and 2008, respectively, and his Ph.D. degree from The University of British Columbia, Vancouver, BC, Canada, in November 2012. 
 He has been listed as a Highly Cited Researcher by Clarivate since 2018.

\textbf{Jinhong Yuan} (Fellow, IEEE) is currently the Head of the School of EE\&T, UNSW. He received the B.E. and Ph.D. degrees in electronics engineering in 1991 and 1997, respectively.  
He served as the IEEE NSW Chapter Chair for the Joint Communications/Signal Processions/Ocean Engineering Chapter from 2011 to 2014 and served as an Associate Editor for IEEE Transactions on Communications from 2012 to 2017 and from 2020 to 2025 and IEEE Transactions on Wireless Communications from 2019 to 2024.
\end{document}